\documentclass{article}%
\usepackage{amsmath}
\usepackage{amsfonts}
\usepackage{amssymb}
\usepackage{graphicx}%
\providecommand{\U}[1]{\protect\rule{.1in}{.1in}}

\begin{document}

\title{Gravitationally induced zero modes of the Faddeev-Popov operator in the
Coulomb gauge for Abelian gauge theories }
\author{Fabrizio Canfora$^{1}$, Alex Giacomini$^{2}$ and Julio Oliva$^{2}$.\\$^{1}${\small \textit{Centro de Estudios Cientificos (CECS), Casilla 1469
Valdivia, Chile.}}\\$^{2}${\small \textit{Instituto de Fisica, Facultad de Ciencias, Universidad
Austral de Chile, Valdivia, Chile.}}\\{\small e-mails:
\textit{canfora@cecs.cl,alexgiacomini@cecs.cl,julio.oliva@docentes.uach.cl}}}
\maketitle

\begin{abstract}
It is shown that on curved backgrounds, the Coulomb gauge Faddeev-Popov
operator can have zero modes even in the abelian case. These zero modes cannot
be eliminated by restricting the path integral over a certain region in the
space of gauge potentials. The conditions for the existence of these zero
modes are studied for static spherically symmetric spacetimes in arbitrary
dimensions. For this class of metrics, the general analytic expression of the
metric components in terms of the zero modes is constructed. Such expression
allows to find the asymptotic behavior of background metrics, which induce
zero modes in the Coulomb gauge, an interesting example being the three
dimensional Anti de-Sitter spacetime. Some of the implications for quantum
field theory on curved spacetimes are discussed.

\end{abstract}

\section{Introduction}

Black hole physics as well as the description of the early universe often
requires the definition of a quantum field theory on a curved background. This
can be particularly subtle when the theory is of the Yang-Mills type. Indeed,
in order to define the classical and quantum dynamics of a gauge theory, it is
necessary to perform a gauge fixing. This is already known to be a non-trivial
issue in flat space-time for non-Abelian gauge theories. The Gribov ambiguity
\cite{Gri78}, whose appearance is closely related to the non-trivial topology
of the space of non-Abelian gauge connections, prevents one from achieving a
global gauge fixing in linear derivative gauges like Lorenz, Coulomb and
Landau gauges\footnote{The mentioned gauge conditions, being linear in the
derivative of the gauge field, define suitable gauge fixings in the Euclidean
path integral framework.
\par
{}}. These copies appear when the gauge potential is large enough, in a sense
that is explained in the following section. Another way to see this problem is
to notice that whenever copies appear, the Faddeev-Popov (FP) operator
acquires zero modes, so that the FP determinant vanishes, and this prevents
one from defining properly the path integral. Since the presence of the copies
in QCD is a non-perturbative phenomenon one can safely compute Feynman
diagrams in linear derivative gauges for the perturbative regime. Indeed, it
was pointed out for the first time in \cite{Gri78}, that the Gribov ambiguity
could also provide a natural explanation for confinement in QCD, once the path
integral is restricted to a region in which the FP operator is positive
definite and therefore, free of zero modes (see e.g. \cite{Zwa96}; other
reviews containing more recent results are \cite{SS05} \cite{EPZ04}
\cite{ILM07}). These ambiguities in the global fixing, of a gauge which is
linear in the derivatives, may also arise in the context of abelian gauge
theories \cite{kelnhofer}, since they can also get a nontrivial structure of
the fibre bundle due, for example, to finite temperature effects \cite{Par88}
\cite{Be09}.

All the cited examples have in common that at least for small enough vector
potentials, the Coulomb or the Euclidean Lorenz gauge fixings are well
defined. The aim of this work is to point out the existence of an infinite
number of zero modes for the FP operator for the Coulomb gauge, on certain
curved space-times, independently on whether the gauge connection is large or
not.\footnote{There are many other results that also show the subtlelties
involved in defining a proper quantization of gauge theories on curved
spacetimes, as for example \cite{TW1}; the classic review of these effects
\ being \cite{BD}.} This is due to the basic fact that on a curved background,
the gauge fixings which are linear in the derivative, are defined in terms of
the covariant derivative instead of a simple partial derivative. Therefore the
equation for the zero modes of the Faddeev-Popov operator depends explicitly
on the metric tensor of the background, through the Christoffell symbols. In
particular in the Coulomb gauge, the equation for the zero modes of the FP
operator is the curved Laplace equation, $\nabla_{i}\nabla^{i}U=0$ where $i$
is an index on the spacelike section, and in a curved space-time, non-singular
solutions to this equation may exist. These solutions may fulfill strong
enough fall-off conditions in order to define a proper (normalizable) gauge
transformation. When this occurs, the restriction of the path integral to some
region of the functional space of vector potential does not help in
eliminating the zero modes, since these appear due to the properties of the background.

Here the appearance of these \textquotedblleft
gravitationally\textquotedblright\ induced zero modes, is studied for static
spherically symmetric backgrounds in arbitrary dimensions. We construct a
static spherically symmetric background, which generates zero modes of the
Coulomb gauge FP operator. We probe also that the three dimensional Anti
de-Sitter (AdS$_{3}$) space-time belongs to this family. This is of special
importance for the AdS/CFT correspondence \cite{maldacena}.

The structure of the paper is the following: in section two, an introduction
to the subtleties of gauge fixing in Yang-Mills theory is given. In section
three, we discuss a set of necessary conditions that a spherically symmetric
space-time should satisfy in order to induce zero modes of the FP operator in
the Coulomb gauge. In section four, a number of simple examples are
considered, including $AdS_{3}$, the three dimensional
Ba\~{n}ados-Teitelboim-Zanelli (BTZ) black hole \cite{BTZ12}, and a four
dimensional wormhole \cite{tw}. Some discussion on possible solutions to the
problem of gauge fixing are finally drawn in the last section.

\section{Gauge-fixing problems in non-Abelian Yang-Mills theory}

In this section a short discussion of the well known gauge fixing problems in
non-Abelian Yang-Mills theory will be presented. These results will suggest
later a very natural and simple solution (yet, the physical consequences of
such a solution are highly non-trivial) to the gravitationally induced gauge
fixing problems analyzed in the next sections.

The appearance of zero modes in the FP operator in non-Abelian Yang-Mills on
flat backgrounds, is related to the well known Gribov problem. In theories
with local symmetries the path integral measure is ill-defined: the action is
invariant under a very large class of local symmetries acting on the
fields,\ so that path-integrating over all the field configurations gives rise
to an overcounting, since many configurations are gauge equivalent. The idea
is then to choose only one field configuration in each equivalence class,
which is achieved by a gauge fixing term. The gauge fixing is in general not
well defined, since when the field amplitude is large enough, the gauge fixing
functional may have the same value when evaluated in gauge equivalent
configurations, violating then the assumption that only one configuration in
each class will fulfill the gauge fixing expression. In gauge theories such a
problem can be avoided, for instance, by using the axial gauge or the
temporal\footnote{Other methods to deal with this issue can be found in the
original reference \cite{Gri78} and, for instance, in the review
\cite{EPZ04}.}, which possesses its own subtleties (see, for instance,
\cite{DeW03}).

The Yang-Mills Lagrangian depends on a Lie algebra valued differential
one-form $A_{\mu}^{a}$ (the most studied case being the one in which the
vector potential is in the adjoint representation of the gauge group) as
follows:
\[
L=-\frac{1}{4g^{2}}trF_{\mu\nu}F^{\mu\nu},\quad\left(  F_{\mu\nu}\right)
^{a}=\left(  \partial_{\mu}A_{\nu}-\partial_{\nu}A_{\mu}+\left[  A_{\mu
},A_{\nu}\right]  \right)  ^{a}.
\]
Such a Lagrangian is invariant under the following (finite) gauge
transformations
\begin{equation}
A_{\mu}\rightarrow A_{\mu}^{\prime}=U^{\dag}A_{\mu}U+U^{\dag}\partial_{\mu
}U,\quad U^{\dag}=U^{-1}, \label{gri0.5}%
\end{equation}
where $U$ is a group-valued scalar\footnote{In this section the $SU(N)$ case
will be considered and the spacetime will be considered flat and Euclidean.}.

Because of the above mentioned invariance, the Lagrangian contains
non-physical degrees of freedom which would implies an overcounting of
configurations in the Euclidean path integral, and for this reason a gauge
fixing is needed. Two of the most common gauge fixings are the Euclidean
Lorenz gauge and the Coulomb gauge, respectively given by
\begin{equation}
\partial_{\mu}A^{\mu}=0;\;\;\;\partial_{i}A^{i}=0 \label{lorentzcoulomd}%
\end{equation}
where $\mu=0,..,3$, while the spatial indices go from $i=1,..,3$. A natural
question then arises: does condition (\ref{lorentzcoulomd}) uniquely fix the
vector potential for each class of configurations which are related by gauge
transformations with suitable fall-off conditions at infinity? As it was shown
by Gribov, the answer is in general negative and many gauge equivalent fields
can satisfy\textit{ the same Coulomb, Euclidean Lorenz or Landau gauge
condition}. As stated above, these gauge fields are related by gauge
transformations which are well-defined, and this will produce an overcounting
of configurations in the path integral approach.

Let us consider the Coulomb gauge. In order for two gauge equivalent fields
$A_{(1)}^{\mu}$ and $A_{(2)}^{\mu}$\ to satisfy the gauge fixing condition, a
group-valued scalar function $U$ has to exist and it should fulfill the
following equations%

\begin{align}
A_{i}^{(2)}  &  =U^{\dag}A_{i}^{(1)}U+U^{\dag}\partial_{i}U\label{gri1}\\
\partial^{i}A_{i}^{(2)}  &  =\partial^{i}A_{i}^{(1)}=0\ . \label{gri2}%
\end{align}

On flat space-time, in the case of non Abelian gauge theories, even when
$A_{(1)}$\ vanishes, there could exist solution of equations (\ref{gri1}) and
(\ref{gri2}), giving rise to $A_{(2)}$ whose spatial components decreases at
spatial infinity as $r^{-\alpha}$ ($r$ being the Euclidean distance and
$\alpha>0$). Anyway these solutions do not decrease fast enough to define a
proper gauge transformation. Even more, a proper gauge transformation must not
only decrease fast enough but must also be non-singular everywhere. However,
if $A_{(1)}$\ is large enough:%

\begin{equation}
\left\Vert A_{(1)}\right\Vert \gtrsim\left\Vert A_{G}\right\Vert ,
\label{critical}%
\end{equation}
where $A_{G}^{\mu}$\ is the critical Gribov\ field and $\left\Vert
.\right\Vert $\ is a suitable norm\footnote{This norm is discussed in details
in the classic papers \cite{DZ89,Va92,Zw82,Zw89}.}, then equations
(\ref{gri1}) and (\ref{gri2}) have smooth solutions in which $A_{(2)}$
decreases rapidly at infinity and therefore, the gauge field is affected by
the Gribov ambiguity. As it is well known, the Gribov ambiguity affects a
proper definition of the path integral, since due to the presence of
non-trivial Gribov copies, the FP operator acquires zero modes and then the FP
determinant vanishes.

In the path integral approach, a possible solution for this ambiguity is to
restrict the integration range in the functional space, to a region in which
the FP operator is positive definite. In such a way the gauge related
configurations would be counted only once (a detailed analysis of this
approach can be found in \cite{DZ89,Va92,Zw82,Zw89}).

On the other hand, in the Abelian case on flat backgrounds, being%
\begin{equation}
U=\exp(i\phi)\ ,
\end{equation}
the condition for the appearance of zero modes of the FP operator in the
Euclidean Lorenz gauge reduces to
\begin{equation}
\square_{E}\phi=0\ , \label{AbelianGri1}%
\end{equation}
while in the Coulomb gauge reduces to:%
\begin{equation}
\Delta\phi=0\ . \label{AbelianGriCou1}%
\end{equation}
Here $\Delta$\ is the three-dimensional Laplace operator and $\square_{E}$\ is
the Euclidean D'Alambert operator, in which the time has been Wick-rotated
$t\rightarrow it\ $.

This means that the gauge field does not enter explicitly in the equation for
the Gribov copies, so that in the Abelian case the existence of the Gribov
ambiguities cannot be eliminated by restricting the path integral to given
region as it occurs in the non-Abelian case. As it is well known, on flat
space-time suitable boundary conditions make invertible both $\square_{E}%
$\ and $\Delta$, or in other words, there are no non-singular solutions with
strong enough fall-off condition of the Laplace equation, avoiding the
appearance of copies in Coulomb and Euclidean Lorenz gauges. However, on
curved space-times, and depending on the exact form of the metric,
non-singular and normalizable copies may exist. This is shown in the next section.

In the Coulomb gauge, suitable boundary conditions (see, for instance,
\cite{DZ89}-\cite{Zw89}; for two detailed reviews see also \cite{SS05}%
\ \cite{EPZ04}) on the gauge transformation are
\begin{align}
&  \phi\underset{r\rightarrow\infty}{\rightarrow}0,\label{normcond1}\\
\mathcal{N}\left(  \phi\right)   &  :=%
{\displaystyle\int_{\Sigma}}
\sqrt{\sigma}d^{3}\Sigma\ \left(  \nabla_{i}\phi\right)  \left(  \nabla
^{i}\phi\right)  <\infty, \label{normcond2}%
\end{align}
where $\sigma$ is the determinant of the induced metric on the spatial section
$\Sigma$. On flat space-times, $\nabla_{i}$ reduces to the partial derivative
$\partial_{i}$ and the indices are raised and lowered with Minkowski metric.
On the other hand on a curved background, the metric $g_{\mu\nu}$ and the
corresponding covariant derivatives have to be used. The above condition
(\ref{normcond2}) on the norm is derived from the requirement that, if before
the gauge transformation one has%
\begin{equation}%
{\displaystyle\int_{\Sigma}}
\sqrt{\sigma}d^{3}\Sigma\ A_{i}A^{i}<\infty,
\end{equation}
then, after the gauge transformation
\begin{equation}
A_{i}\rightarrow\left(  A^{\phi}\right)  _{i}=A_{i}+\partial_{i}\phi,
\end{equation}
one should also have%
\begin{equation}%
{\displaystyle\int_{\Sigma}}
\sqrt{\sigma}d^{3}\Sigma\ \left(  A^{\phi}\right)  _{i}\left(  A^{\phi
}\right)  ^{i}<\infty.
\end{equation}
In other words, one should ask for $\nabla_{i}\phi$ to satisfies the same
fall-off conditions that one requires for $A$.\ Unlike the non-Abelian case,
to avoid zero modes of the FP operator in the Abelian case, it is enough to
consider non-singular fields decreasing at infinity and to notice that, on
flat backgrounds, the Laplace operator does not posses everywhere smooth and
non-singular solutions, decreasing fast enough at infinity.

On the other hand, when defining the path integral on a curved background in
the Coulomb gauge for an Abelian gauge theory, it explicitly appears the FP
determinant $J$:%
\begin{equation}
J=\det\nabla_{i}\nabla^{i}\ , \label{shun-pei}%
\end{equation}
which represents the Jacobian of the change of coordinates in the functional
space. Therefore, if there exist non-trivial and everywhere smooth regular
solutions of the equation%
\begin{equation}
\nabla_{i}\nabla^{i}\phi=0\ , \label{gricoucurved1}%
\end{equation}
then the determinant in Eq. (\ref{shun-pei}) vanishes and the path integral in
the Coulomb gauge cannot be defined properly\footnote{In flat spacetimes the
operators $\nabla_{i}$ are nothing but the usual partial derivatives
($\nabla_{i}=\partial_{i}$) so that, as it has been already discussed, the FP
determinant $J=\det\partial_{i}\partial^{i}$ does not vanish.}. On curved
backgrounds, the operator $\nabla_{i}$ is the covariant derivative along the
spatial directions on which one would like to define a path integral
quantization of an Abelian gauge theory. Therefore, in some interesting cases
non-trivial smooth everywhere regular solutions\ of equation
(\ref{gricoucurved1}) do indeed exist as it will be shown in the next
sections. One could still try to insist in factoring out the zero modes of the
FP operator $\nabla_{i}\nabla^{i}$ when computing the determinant: this
procedure seems to be reasonable when the operator just has one, or few zero
modes. Nevertheless in what follows, we will construct examples in which the
FP operator $\nabla_{i}\nabla^{i}$ has infinitely many zero modes and, in
these cases, the factorizing out procedure appears quite unnatural. It is
worth to stress that one cannot exclude the existence of zero modes by just
imposing stronger asymptotic fall-off conditions on the gauge potential since
this would correspond to arbitrarily \textquotedblleft
mutilate\textquotedblright\ the solution space of the gauge field
\cite{reggeteitelboim}.

\section{Coulomb gauge in static spherically symmetric space-times}

Let us consider a static spherically symmetric metric in $d$ dimensions of the
form%
\begin{equation}
ds_{d}^{2}=-f\left(  r\right)  dt^{2}+\frac{dr^{2}}{f\left(  r\right)  }%
+r^{2}d\Omega_{d-2}^{2}\ , \label{september1}%
\end{equation}
where the spatial section is the warped product of the real line and a
$d-2$-dimensional sphere $S^{d-2}$.

The Coulomb gauge condition reads%
\begin{equation}
\nabla^{i}A_{i}=0\ ,
\end{equation}
where $\nabla^{i}$\ is the covariant derivative along the spatial directions
($i=1$, $2$,$...$, $d-1$) of the metric (\ref{september1}), and the $A_{i}$
are the spatial components of the Abelian vector potential. Therefore, the
equation which determines the appearance of a zero mode $\phi$ of the FP
operator in the Coulomb gauge simply reduces to the Laplace equation on the
spatial section of (\ref{september1}), i.e.:%
\begin{equation}
\nabla^{i}\nabla_{i}\phi=0\ . \label{september2}%
\end{equation}
Let us now decompose the gauge parameter $\phi$\ as follows
\begin{equation}
\phi=F\left(  r\right)  Y\left(  S^{d-2}\right)  \ , \label{separation}%
\end{equation}
where $Y\left(  S^{d-2}\right)  $ is a generalized spherical harmonic on the
$d-2$-dimensional sphere%
\begin{equation}
\nabla_{S^{d-2}}^{2}Y=-QY\ , \label{september3}%
\end{equation}
with $Q$ being the corresponding eigenvalue \thinspace$l\left(  l+d-3\right)
$. Equation (\ref{september2}) reduces to%
\begin{equation}
2fF^{\prime\prime}+\left(  2\left(  d-2\right)  f+rf^{\prime}\right)
r^{-1}F^{\prime}-2Qr^{-2}F=0\ . \label{september4}%
\end{equation}
where prime denotes derivative with respect to the argument.

Interestingly enough, Eq. (\ref{september4}) allows to integrate the metric
function $f\left(  r\right)  $ in terms of the radial part $F(r)$ of the zero
mode:%
\begin{equation}
f(r)=\frac{r^{-2d+4}\left(  \int^{r}Q\left(  F^{2}\left(  x\right)  \right)
^{\prime}x^{2d-6}\ dx+I_{1}\right)  }{\left(  F^{\prime}\right)  ^{2}}\ ,
\label{september5}%
\end{equation}
where $I_{1}$ is an integration constant. This simple relation allows to draw
some general conclusions about how a static spherically symmetric
gravitational field has to behave, in order to induce zero modes of the FP
operator in the Coulomb gauge. In other words, assuming that $\phi$ is an
everywhere regular smooth gauge transformation, it is possible to determine
the asymptotic behavior of the metric for $r\rightarrow\infty$. It is worth
stressing here the close similarity of the present procedure which allows one
to express the gravitational field in terms of the zero modes of the FP
operator, with the well known technique in supersymmetric quantum mechanics
which allows to write the quantum Hamiltonian in terms of the corresponding
zero mode (see, in particular, \cite{Witten} \cite{Gozzi} \cite{Zanelli}).

\section{Some comments on the generic behavior of backgrounds inducing zero
modes of the FP operator}

In this section we will concentrate on the asymptotic behavior of the metric
function $f(r)$ obtained from Eq. (\ref{september5}) at infinity, i.e. when
$r\rightarrow\infty$.

Since the hypothesis is that the zero mode $\phi=F\left(  r\right)  Y\left(
S^{d-2}\right)  $ is everywhere regular and smooth, let us assume that for
$r\rightarrow\infty$, the behavior of $F(r)$ is%
\begin{equation}
F(r)\underset{r\rightarrow\infty}{\approx}c_{0}+\frac{c_{1}}{r^{n}%
}+...,\text{\ \ \ \ }n>0,\ \ \ \ \ \ c_{0}\neq0,\ \ \ \ \ c_{1}\neq0,
\label{september5.95}%
\end{equation}
where ($...$) denote subleading terms and $c_{0}$, $c_{1}$ are constants (the
analysis in the cases in which $c_{0}$ and/or $c_{1}$ vanish/es follows the
same lines). Then%
\begin{equation}
F^{2}\underset{r\rightarrow\infty}{\approx}\left(  c_{0}\right)  ^{2}%
+\frac{2c_{0}c_{1}}{r^{n}}+...,\ \ \ \ F^{\prime}\underset{r\rightarrow\infty
}{\approx}-\frac{nc_{1}}{r^{n+1}}+...\ .
\end{equation}
Therefore, in the asymptotic region $r\rightarrow\infty$ using
(\ref{september5}) when $c_{0}\neq0$, one obtains the following expression for
$f\left(  r\right)  $:%
\begin{equation}
f(r)\underset{r\rightarrow\infty}{\approx}\frac{I_{1}}{c_{1}^{2}n^{2}%
}r^{6+2n-2d}-\frac{2Qc_{0}}{nc_{1}\left(  2d-n-6\right)  }r^{n}-\frac
{Q}{n\left(  d-n-3\right)  }\ ,\ \ \ \ \ \ \ \ n\neq2d-6\text{ and }n\neq
d-3\ . \label{fc0no0}%
\end{equation}
Then this is the generic expression for the metric function $f\left(
r\right)  $ when the zero mode of the FP operator takes the form
(\ref{september5.95})\footnote{For the special values $n=2\left(  d-3\right)
$ or $n=d-3$, logarithmic branches may appear, and the analisys follows along
the same lines.}.

On the other hand, when $c_{0}=0$ in Eq. (\ref{september5.95}) one obtains
\begin{equation}
f(r)\underset{r\rightarrow\infty}{\approx}\frac{I_{1}}{c_{1}^{2}n^{2}%
}r^{6+2n-2d}-\frac{Q}{n\left(  d-n-3\right)  }\ ,\ \ \ \ \ \ \ \ n\neq d-3\ .
\label{september6.5}%
\end{equation}

The requirement that on the space-time (\ref{september1}), the norm
(\ref{normcond2}) of the gauge transformation $\phi=F\left(  r\right)
Y\left(  S^{d-2}\right)  $ has to remain finite reduces to%
\begin{equation}
\mathcal{N}\left(  \phi\right)  :=\left[  \int dr\ r^{d-2}f^{1/2}\left(
F^{\prime}\right)  ^{2}+r^{d-4}f^{-1/2}F^{2}Q\right]  \left[  \int_{S^{d-2}%
}Y^{\ast}Y\sqrt{\Omega_{d-2}}d^{d-2}x\right]  <\infty\ . \label{normsplitted}%
\end{equation}
The angular integral is finite and so one needs to take care only of the
radial integral. When $c_{0}\neq0$, replacing in the expressions
(\ref{fc0no0}) and (\ref{september5.95}) in the norm (\ref{normsplitted}) one
concludes that the convergence of the latter is guaranteed when:%
\begin{align}
n  &  >2\left(  d-3\right)  \text{ or}\label{relax1}\\
n  &  \leq2\left(  d-3\right)  \text{ and }Q=0\text{ with }I_{1}\neq0\ .
\label{relax2}%
\end{align}
For $c_{0}=0$, the expression for the metric function $f\left(  r\right)  $ is
given by equation (\ref{september6.5}) and the norm (\ref{normsplitted})
converges if%
\begin{equation}
n>\frac{d-3}{2}\ . \label{nlowerc0igual0}%
\end{equation}

The case with $d=3$ must be analyzed separately: the radial part $F\left(
r\right)  $ of the zero mode and the metric function $f\left(  r\right)  $ are
given by (\ref{september5.95}) and (\ref{september5}) respectively. In this
case, the norm (\ref{normsplitted}) always converges at infinity. In the next
section, some explicit three dimensional space-times which induce zero modes
of the FP operator will be constructed.

Based on the above discussion, it is possible to draw some interesting conclusions.

If the space-time is asymptotically locally flat, $f(r)$ in (\ref{september1})
should behave as follows%
\begin{equation}
f\underset{r\rightarrow\infty}{\approx}1+\frac{c}{r^{p}}%
+...\ ,\ \ \ \ \ p>0\ , \label{schw}%
\end{equation}
$c$ and $p$ being constants. Thus, in such cases, induced zero modes of the FP
operator in the Coulomb gauge could appear only when $c_{0}=0$ and $n<d-3$
(see Eq. (\ref{september6.5})). However when $c_{0}=0$, the finiteness of the
norm $\mathcal{N}\left(  \phi\right)  $ in (\ref{normsplitted}) implies
equation (\ref{nlowerc0igual0}), and consequently\footnote{Note that again
this bounds suggest that the case $d=3$ should be analyzed separately.} in
this case zero modes of the FP operator would exist provided
\begin{equation}
\left(  d-3\right)  /2<n<d-3\ . \label{blabla}%
\end{equation}
From this analysis it is also clear that, an asymptotically Schwarzschild
space-time for which the asymptotic behavior of $f\left(  r\right)  $ is given
by (\ref{schw}) with $p=d-3$, will not generate zero modes of the FP operator.
This is due to the fact that comparing (\ref{schw}) and (\ref{september6.5})
one obtains $n=\left(  d-3\right)  /2$, giving rise to a non-normalizable zero
mode (see equation (\ref{blabla})).

\bigskip

Another interesting asymptotic behavior is defined by the $AdS$ space-times.
In $AdS$ and asymptotically $AdS$ metrics, fields propagating in the bulk have
been conjectured to be dual to a conformal field theory in the boundary
\cite{maldacena}: the so-called $AdS/CFT$\ correspondence opens the remarkable
possibility to explore the non-perturbative regime of supersymmetric
Yang-Mills theories by performing semiclassical computations in the bulk of
asymptotically $AdS$ background (see also\textbf{ }\cite{magoo}). Furthermore,
this idea has been also used to give a consistent microscopic interpretation
of the black hole entropy. In particular, unlike the higher dimensional cases,
in $2+1$ dimensions the algebra of the asymptotic symmetries of an
asymptotically AdS space-time is enlarged to two copies of the infinite
dimensional Virasoro algebra, whose central charge can be used to compute the
entropy of the BTZ black hole \cite{BTZ12} by means of the Cardy formula
\cite{cardy}, \cite{strom}. These reasons make $AdS_{3}$ as well as the BTZ
black hole two of the most important curved geometries in high energy physics.
In the next section, it will be show that the problem to find normalizable
zero modes of the FP operator both on $AdS_{3}$ and on the BTZ black hole can
be solved in a quite elegant and simple way. It will be also shown how to
construct normalizable zero modes of the FP operator on an interesting four
dimensional wormhole space-time.

\section{Explicit examples of backgrounds generating zero modes of the FP
operator}

\subsection{$AdS_{3}$ space-time}

As mentioned above, the three dimensional $AdS$ space-time is of special
interest for high energy physics in view of the $AdS/CFT$ correspondence
\cite{maldacena}. The $AdS_{3}$ metric is given by%
\begin{equation}
ds^{2}=-\left(  \frac{r^{2}}{l^{2}}+1\right)  dt^{2}+\frac{dr^{2}}{\frac
{r^{2}}{l^{2}}+1}+r^{2}d\psi^{2}\ , \label{ads3}%
\end{equation}
where $-\infty<t<+\infty$, the radial coordinate runs in the range $0\leq
r<+\infty$, and the angular coordinates fulfills $0\leq\psi\leq2\pi$, $0$
being identified with $2\pi$. This space-time is of constant and negative
curvature. It solves the Einstein equations with a negative cosmological
constant $\Lambda=-\frac{1}{l^{2}}$, where $l$ is the AdS radius. Assuming the
separation (\ref{separation}) for the zero mode, where now the spherical
harmonics are%
\begin{equation}
Y\left(  S^{1}\right)  \sim e^{im\psi}\ , \label{YS1}%
\end{equation}
with $m$ an integer, the equation (\ref{september2}) reduces to%
\begin{equation}
2r^{2}fF^{\prime\prime}+r\left(  2f+rf^{\prime}\right)  F^{\prime}%
-2m^{2}F=0\ ,
\end{equation}
where $f\left(  r\right)  =\frac{r^{2}}{l^{2}}+1$. For positive $m$, the
regular solution at the origin reads%
\begin{equation}
F\left(  r\right)  =C\frac{\left(  r/l\right)  ^{m}}{\left(  1+\sqrt
{\frac{r^{2}}{l^{2}}+1}\right)  ^{m}}\ ,
\end{equation}
$C$ being an integration constant\footnote{For negative $m$ the solution which
is regular at the origin is preciselly the one we discarded in the case with
$m$ positive, and the solution takes the same expression.}. This solution is
regular everywhere, and when $r$ goes to infinity $F\left(  r\right)  $ tends
to the constant $C$. The norm (\ref{normsplitted}) in this case reduces to%
\begin{equation}
\mathcal{N}\left(  \phi\right)  =4\pi C^{2}m^{2}\int_{0}^{\infty}%
\frac{lr^{2m-1}\left(  r^{2}+2l^{2}+2l\sqrt{r^{2}+l^{2}}\right)  }{\sqrt
{r^{2}+l^{2}}\left(  l+\sqrt{r^{2}+l^{2}}\right)  ^{2m+2}}dr\ .
\end{equation}
This integral converges, since the integrand (denoted by $H_{AdS_{3}}$) is
smooth, it goes to zero at the origin and also goes to zero when $r$ goes to
infinity as%
\begin{equation}
H_{AdS_{3}}\underset{r\rightarrow\infty}{\sim}\frac{1}{r^{2}}+O\left(
1/r^{3}\right)  \ .
\end{equation}
Thus $AdS_{3}$ generates infinite zero modes of the FP operator. This result
is particularly relevant for the AdS/CFT conjecture, and it would be nice to
further analyze the consequences of the existence of these zero modes from the
point of view of the dual boundary CFT.

\subsection{Zero modes of the FP operator on the BTZ black hole}

The metric for the static BTZ black hole is given by \cite{BTZ12}%
\begin{equation}
ds^{2}=-\left(  \frac{r^{2}}{l^{2}}-\mu\right)  dt^{2}+\frac{dr^{2}}%
{\frac{r^{2}}{l^{2}}-\mu}+r^{2}d\psi^{2}\ , \label{btz}%
\end{equation}
with $0<\psi\leq2\pi$, and $0$ identified with $2\pi$. This space-time
describes an asymptotically AdS black hole whose mass is proportional to the
parameter $\mu$. The corresponding event horizon is located at $r=r_{+}%
=l\sqrt{\mu}$. The metric (\ref{btz}) can be also obtained by an
identification of the three dimensional $AdS_{3}$ space-time (\ref{ads3})
\cite{bhtz}. For our purposes, we will consider only the exterior part of the
black hole (\ref{btz}):%
\begin{equation}
l\sqrt{\mu}<r<+\infty\ ,
\end{equation}
since the Euclidean continuation of the black hole metric (which is necessary
to achieve a proper definition of black hole thermodynamics) covers only the
region outside of the event horizon. It is useful to perform the change of
coordinates
\begin{equation}
r=\frac{l\sqrt{\mu}}{\sqrt{1-x^{2}}}\ ,
\end{equation}
with $0<x<1$, the surfaces $x=0$ and $x=1$ being the horizon and spatial
infinity respectively. With this change of coordinates the BTZ metric
(\ref{btz}) reduces to%
\begin{equation}
ds^{2}=-\frac{\mu x^{2}}{1-x^{2}}dt^{2}+\frac{l^{2}dx^{2}}{\left(
1-x^{2}\right)  ^{2}}+\frac{l^{2}\mu}{1-x^{2}}d\psi^{2}\ .
\end{equation}
Assuming that the angular part of the zero mode is given by (\ref{YS1}), the
equation (\ref{september2}) for the corresponding radial part reads%
\begin{equation}
\mu\left(  1-x^{2}\right)  F^{\prime\prime}\left(  x\right)  -\mu xF^{\prime
}\left(  x\right)  -m^{2}F\left(  x\right)  =0\ .
\end{equation}
The solution of the above equation is given by%
\begin{equation}
F\left(  x\right)  =\alpha_{1}\left(  x+\sqrt{x^{2}-1}\right)  ^{\frac
{im}{\sqrt{\mu}}}+\alpha_{2}\left(  x+\sqrt{x^{2}-1}\right)  ^{-\frac
{im}{\sqrt{\mu}}}\ ,
\end{equation}
$\alpha_{1}$ and $\alpha_{2}$ being integration constants. The norm in this
case reduces to%
\begin{equation}
\mathcal{N}\left(  \phi\right)  =4\pi^{2}\alpha_{1}\alpha_{2}\mu^{-1/2}%
m^{2}\ , \label{normBTZ}%
\end{equation}
being finite for any value of\footnote{From equation (\ref{normBTZ}), one may
think that the massless BTZ black hole, in which $\mu\rightarrow0$, does not
possess any normalizable zero mode. Nevertheless analyzing the problem from
scratch in such a geometry one is able to show that also in this case
infinitely many normalizable zero modes exist.} $m$.

\subsection{FP zero modes in a four dimensional wormhole space-time}

Wormholes are some of the most fascinating curved backgrounds since they
posses two (or more) asymptotic regions connected by one (or more) throat(s).
Let us consider the four dimensional spherically symmetric wormhole metric%
\begin{equation}
ds^{2}=-dt^{2}+d\rho^{2}+\cosh^{2}\rho d\Omega_{2}^{2}\ , \label{wormh}%
\end{equation}
where $d\Omega_{2}^{2}$ is the line element for the two sphere. The two
asymptotic regions posses geometries locally equivalent to $\mathcal{R}\times
H_{3}$ ($H_{3}$ being a negative three dimensional constant curvature
manifold) and are located at $\rho\rightarrow\pm\infty$. These boundaries are
connected by the throat being located at $\rho=0$. As shown in reference
\cite{tw}, this space-time is a vacuum solution of Weyl conformal gravity in
four dimensions\footnote{This wormhole have been extended to diverse
dimensions within certain gravity theories in \cite{ewt1}-\cite{ewt4}.}: in
this theory the field equations reduce to the vanishing of the Bach tensor
(for some relevant references see also \cite{CW1}-\cite{CW3}). It is
interesting to note that wormhole geometries in four dimension can also arise
in General Relativity with matter sources localized around the throat and
violating standard energy conditions (this type of sources can arise from
quantum fluctuations \cite{VisserBook}). The metric (\ref{wormh}) provides a
simple background generating zero modes of the FP operator.

After the change of coordinates $\rho=\tanh^{-1}x$, the equation for the
(radial part of the) FP zero mode in this case acquires a very simple form%
\begin{equation}
\left(  1-x^{2}\right)  F^{\prime\prime}\left(  x\right)  -QF\left(  x\right)
=0\ ,
\end{equation}
which is integrated in terms of hypergeometric functions. The norm of the zero
mode is finite at both asymptotic regions $x\rightarrow\pm1$ only when $Q=0$
(the "S-wave"). In this case, the normalizable zero mode is given by%
\begin{equation}
F\left(  x\right)  =\beta\left(  1-x\right)  \ ,
\end{equation}
$\beta$ being an integration constant. In the original coordinates system the
FP zero mode reads%
\begin{equation}
\phi=\frac{\beta}{\sqrt{4\pi}}\left(  1-x\right)  =\frac{\beta}{\sqrt{4\pi}%
}\left(  1-\tanh\rho\right)  \ ,
\end{equation}
whose norm reduces to%
\begin{equation}
\mathcal{N}\left(  \phi\right)  =2\beta^{2}\mathbf{\ .}%
\end{equation}

\section{Possible solutions to gravitationally-induced gauge fixing problems
and further comments}

In the previous sections, we have discussed the asymptotic behavior of static
spherically symmetric curved backgrounds which induces zero modes of the FP
operator in the Coulomb gauge and we have also constructed interesting
concrete examples. The aim of the previous sections was first to show that
this phenomenon is quite generic and secondly that also curved backgrounds of
great theoretical interest are affected by this problem. The question of how
to solve this problem, then naturally arises. A possible point of view is that
this is not a problem at all since in the present case, the metric-dependent
FP determinant $\det\left(  -\nabla_{i}\nabla^{i}\right)  $ does not depend on
the gauge field and, therefore, it factorizes out from the path integral on
the gauge field:%
\begin{equation}
Z=\int DA_{\mu}\det\left(  -\nabla_{i}\nabla^{i}\right)  \exp\left(
S_{QED}\right)  =\det\left(  -\nabla_{i}\nabla^{i}\right)  \int DA_{\mu}%
\exp\left(  S_{QED}\right)  \label{bawa1}%
\end{equation}
where $S_{QED}$\ is the standard QED action. Nevertheless, this point of view
is not very appealing due to the following reasons.

The first and main reason arises when one considers the semiclassical
quantization of gravity\footnote{We will not dwell here on which is the final
theory of quantum gravity, nevertheless, it is worth to note that the
gravitational action in three dimensions is renormalizable (see e.g.
\cite{revisited}).}; whatever the final theory of quantum gravity is, a
semiclassical regime in which the weakly coupled quantum fluctuations of
gravity can be treated perturbatively with the usual method of QFT has to
exist. In this regime, in which the quantum gravitational effects are not
strong (as, for instance, in the first stages of the Hawking evaporation
process), one should treat the graviton ($h_{\mu\nu}$) as a quantum
fluctuation propagating on a classical gravitational background. This makes
compulsory that, when dealing with QED on curved backgrounds, one should
consider the quantum fluctuations of the graviton as well. This point of view
is mandatory when one analyzes, for instance, perturbative regime of
supergravity theories in which the graviton $h_{\mu\nu}$ and the photon
$A_{\mu}$ are in the same supermultiplet or in the already mentioned black
hole evaporation process. Obviously, in this situation, the FP operator does
not factor out from the path integral (on both $A_{\mu}$ and $h_{\mu\nu}$)
anymore. The second reason is that, even in the context of QED on a fixed
curved background when Eq. (\ref{bawa1}) holds, the proper definition of the
FP determinant may be problematic. Indeed, some regularization procedure is
needed otherwise the path integral would be just multiplied by zero.
Furthermore, already at classical level the proper definition of the canonical
formalism through the Dirac brackets would be ill defined since the inverse of
the spacelike Laplacian appears in the denominator of the right hand side of
the Dirac Brackets \cite{HRT}. The usual procedure to achieve this for the
Coulomb gauge on flat space-time would be to regularize the FP determinant by
considering the functional space of functions orthogonal to the zero mode
(which on flat space is the constant function). Such a procedure is viable
since it does not imply any unphysical mutilation of the space of solutions.
On the other hand, if the FP operator admits non-trivial zero modes,\ this
procedure is quite subtle since by restricting the functional space to
functions orthogonal to the non-trivial zero modes one could discard a
physical part of the functional space. Careless mutilations of the space of
solutions are known to lead to unphysical results: the soundest point of view
is to take as physical boundary conditions the ones that lead to a well
defined variational principle \cite{reggeteitelboim}. It seems that this
criterion is not enough to eliminate from the functional space the zero modes
which have been presented here since they satisfy all the required boundary
conditions. Eventually, when a classical background induces an infinite number
of non-trivial zero modes of the FP operator it appears that a suitable
definition of the path integral along these lines would not be possible at all.

The first and most obvious solution to this problem would be to follow the
same procedure as in non-Abelian gauge theories i.e. to restrict the path
integral to a copy free region. In the context of the present paper this means
to exclude spacetimes which induce zero modes of the Coulomb gauge FP operator
from the path integral\footnote{As it is well known, if the Coulomb gauge is
pathological, usually other covariant gauges manifest some kinds of
pathologies as well (see, for instance, \cite{DeW03} \cite{singer}).}. This
point of view is strongly supported by the need to take into account the
effects of gravitational fluctuations in many situations of interest (such as
perturbative quantization of supergravity or the analysis of the Hawking
process) as it has been explained in the discussion above. From the point of
view of Quantum Field Theory, this does not seem to produce any conceptual
difficulty as it has been explained in section 2. Nevertheless from the point
of view of gravitational physics the need to declare, for instance, the
$AdS_{3}$ and BTZ spacetimes unphysical is extremely problematic. This is due
to the fact that one of the most powerful tools for making predictions both on
a possible quantum theory of gravity and on nonperturbative regime in
Yang-Mills theory is given by the AdS/CFT correspondence where the fields
living in an AdS spacetime (including Abelian fields) can be related to a
conformal field theory on the boundary. In the context of black hole physics,
the case of a three dimensional bulk spacetime turns out to be especially
important. The $AdS_{3}/CFT_{2}$ correspondence is one of the most used tools
to describe the gravitational microstates responsible for the black hole
entropy. Therefore, the solution to declare, for instance, the three
dimensional $AdS$ space-time and the BTZ black hole as unphysical would have
highly non-trivial consequences. This issue is worth to be further investigated.

The second and more ambitious solution to gravitationally induced gauge fixing
problems would be to formulate from the very beginning QED in terms of gauge
invariant variables (such as the Wilson loops) which are not affected by gauge
fixing problems. Even if this would be the most elegant possibility, it is not
very practical yet since, up to now, explicit non-trivial computations in
field theory in terms of Wilson loops have been performed only in topological
field theories.

\section{Acknowledgments}

We thank Andr\'{e}s Anabal\'{o}n and Silvio Sorella for enlightening comments.
We would also like to thank the referees of PRD for very useful comments. This
work is supported by Fondecyt grants 11080056 and 11090281, by UACh-DID grant
S-2009-57, and by the Conicyt grant \textquotedblleft Southern Theoretical
Physics Laboratory\textquotedblright\ ACT-91. F. C. is also supported by
PROYECTO INSERCI\'{O}N CONICYT 79090034 and by the Agenzia Spaziale Italiana
(ASI). The Centro de Estudios Cient\'{\i}ficos (CECS) is funded by the Chilean
Government through the Millennium Science Initiative and the Centers of
Excellence Base Financing Program of Conicyt. CECS is also supported by a
group of private companies which at present includes Antofagasta Minerals,
Arauco, Empresas CMPC, Indura, Naviera Ultragas and Telef\'{o}nica del Sur.
CIN is funded by Conicyt and the Gobierno Regional de Los R\'{\i}os.

\end{document}